\documentclass[a4paper,11pt]{article}
\pdfoutput=1 

\usepackage{jcappub} 

\usepackage[T1]{fontenc} 
\usepackage{verbatim}

\usepackage{geometry}                
\usepackage{graphicx}
\usepackage{caption}
\usepackage{subcaption}

\usepackage{amssymb}
\usepackage{amsmath}
\usepackage{epstopdf}
\usepackage{array}
\def\be{\begin{equation}} 
\def\ee{\end{equation}} 
\def\bea{\begin{eqnarray}} 
\def\eea{\end{eqnarray}}

\def\knr{k_{\rm nr}}

\usepackage[numbers]{natbib}
\usepackage[table,usenames,dvipsnames]{xcolor}

\title{Neutrino Masses, Scale-Dependent Growth, and Redshift-Space Distortions}

\author[a,b]{Oscar F. Hern\'andez,}

\affiliation[a]{Marianopolis College, \\ 4873 Westmount Ave.,Westmount, QC H3Y 1X9, Canada}
\affiliation[b]{ 
Department of Physics, McGill University, \\
Montr\'eal, QC, H3A 2T8, Canada}

\emailAdd{oscarh@physics.mcgill.ca}
\date{\today}                      

\abstract{ Massive neutrinos leave a unique signature in the large scale clustering of matter.
We investigate the wavenumber dependence of the growth factor arising from neutrino masses and use a Fisher analysis to determine the aspects of a galaxy survey needed to measure this scale dependence. 
}

\arxivnumber{}

\begin{document}
\maketitle

\section{Introduction}
\label{sec:intro}

The neutrino oscillations confirmed by solar, atmospheric, reactor and accelerator neutrino experiments indicate that the Standard Model neutrinos have non-zero mass (see for example \cite{2014PhRvD..89i3018C,2014JHEP...11..052G,2015JHEP...09..200B}).
Neutrino oscillation experiments constrain mass squared differences between neutrino mass states, but  this type of data is unable to determine alone the absolute neutrino mass scale. Furthermore, with current measurements, different arrangements of the neutrino mass states are allowed for the ordering of the neutrino masses, namely the normal hierarchy (NH) or the inverted hierarchy (IH).

By being sensitive to the gravitational effect of the sum of all neutrino mass states, cosmological observables provide an alternative way of characterizing neutrino masses that is complementary to oscillation experiments (see \cite{2015APh....63...66A} for a review).  Current constraints on the sum of neutrino masses $\sum m_{\nu}$ are approaching the critical 0.1 eV level, below which an inverted neutrino mass hierarchy is excluded  (\cite{2015arXiv151105983C,2015JCAP...11..011P,2015arXiv150201589P}). 
Massive neutrinos are expected to have an effect on the growth of large scale structure in the the universe.  
Massive neutrinos shift matter-radiation equality and reduce cold dark matter (CDM) fluctuations during matter domination (\cite{1996PhRvD..54.1332J,1998ApJ...498..497H}). These effects decrease small-scale perturbations and imply that neutrino mass information can be obtained from the matter power spectrum $P_m(k,z)$.

Many previous studies (\cite{1998PhRvL..80.5255H}, see \cite{2006PhR...429..307L} for a review) 
have investigated $P_m(k,z)$ at large $k\gg\knr$ where analytic approximations exist. The wavenumber $\knr$ corresponds to the horizon
size when the thermal velocities of neutrinos become nonrelativistic: 
$ \knr\approx0.02 (\Omega _m)^{1/2} \sqrt{m_{\nu }\over 1 {\rm eV}}~{h\over \rm Mpc}$.
The high thermal velocities are sufficient to suppress structure formation on small scales, but as the neutrino momenta redshift
away it becomes possible for neutrinos to participate in cosmic growth of structure on larger scales. This leads to a distinct
scale dependence to structure formation that changes with time~\cite{1983ApJ...274..443B}.
This time-dependent scale-dependence is a unique signature of
a particle with a high thermal velocity, and there is a clear prediction for both the amplitude and shape of this effect for 
any particular set of neutrino masses. 

There are several ways that this time-dependent, scale-dependent effect is measurable. 
One approach consists in studying the time evolution of the matter clustering at a given scale. 
Given a measurement of the matter power spectrum at early times (as provided
by the cosmic microwave background, for example),
one could measure the power spectrum at late times to determine exactly
how much power has been suppressed. 
This could be difficult to separate from other effects which change the growth
rate of structure, such as dark energy, but is a sensitive probe of neutrino 
masses (e.g. \cite{2011MNRAS.415.2876B})

Another approach corresponds to measuring the scale dependence of the matter power spectrum at a fixed
time to probe the differential power at large and small scales coming from the integrated difference in growth.
While this could be difficult to separate from a scale dependent bias, it has been shown that scale-dependent
bias is also an additional probe of neutrino masses \cite{2016arXiv160208108L}. 
Performing this measurement at several cosmic times, however, would allow 
a measurement of the time dependent scale-dependence of the
growth of structure. 

In this paper, we investigate a method to measure the dynamics of structure formation that captures the essential aspect of
the rate at which structure grows being dependent on scale: redshift space distortions \cite{1987MNRAS.227....1K}. Measurements of large scale structure
typically use cosmological redshift as a proxy for the line-of-sight distance. Peculiar velocities from the growth of structure lead
to a distortion in this mapping that is simply related to the rate at which large scale structure density contrasts are growing. 
This is a measurement of the growth rate of structure, commonly defined as
\be\label{f}
f=d(\ln D_1)/d\ln a   \ ,
\ee
where $D_1$ is the amplitude of density perturbations and $a$ is the cosmological scale factor.
The free-streaming of neutrinos will slightly suppress this rate on small scales
as compared to large scales, making $f$ a function of both $k$ and $z$. While this is a relatively small effect, it is a direct measurement of the effect of neutrinos on the
growth of large scale structure and would be an extremely robust model-independent validation of a cosmological neutrino mass measurement.

In this paper, we ask how well a galaxy power spectrum measurement 
can determine $f$ and its scale dependence,
the unique signature of a massive (but still relatively light) particle. We approach this question in section~\ref{surveys} through a Fisher matrix analysis and determine the size of the galaxy survey needed to measure the scale 
dependence of $f$. This will depend on redshift, as both the available volume and the scale dependence of $f$ are redshift-dependent. Our results are summarized by figure~\ref{VolfLH} which is presented and discussed in section~\ref{results}.

The growth of structure as a cosmological probe contains a wealth of information on dark matter, dark energy and possible extensions (see  \cite{2015APh....63...23H} for a review). 
The work of \cite{2016arXiv160208108L} studies the scale-dependent, time-dependent effect of massive 
neutrinos of the growth of structures, including galaxy bias and including the effect of using
multiple tracers of large scale structure as a way to improve precision. 
Our work differs in that we restrict ourselves to the case of a single tracer and focus on the detectability of the scale dependence of $f(k,z)$ at different redshifts in order to design future surveys with this goal in mind.  
We refer the interested reader to \cite{2016arXiv160208108L} for further considerations on the scale dependent effect of massive neutrinos on the halo bias and its impact on the total neutrino mass measurement forecasted errors.

\begin{figure}[htbp]
\includegraphics[height=6cm]{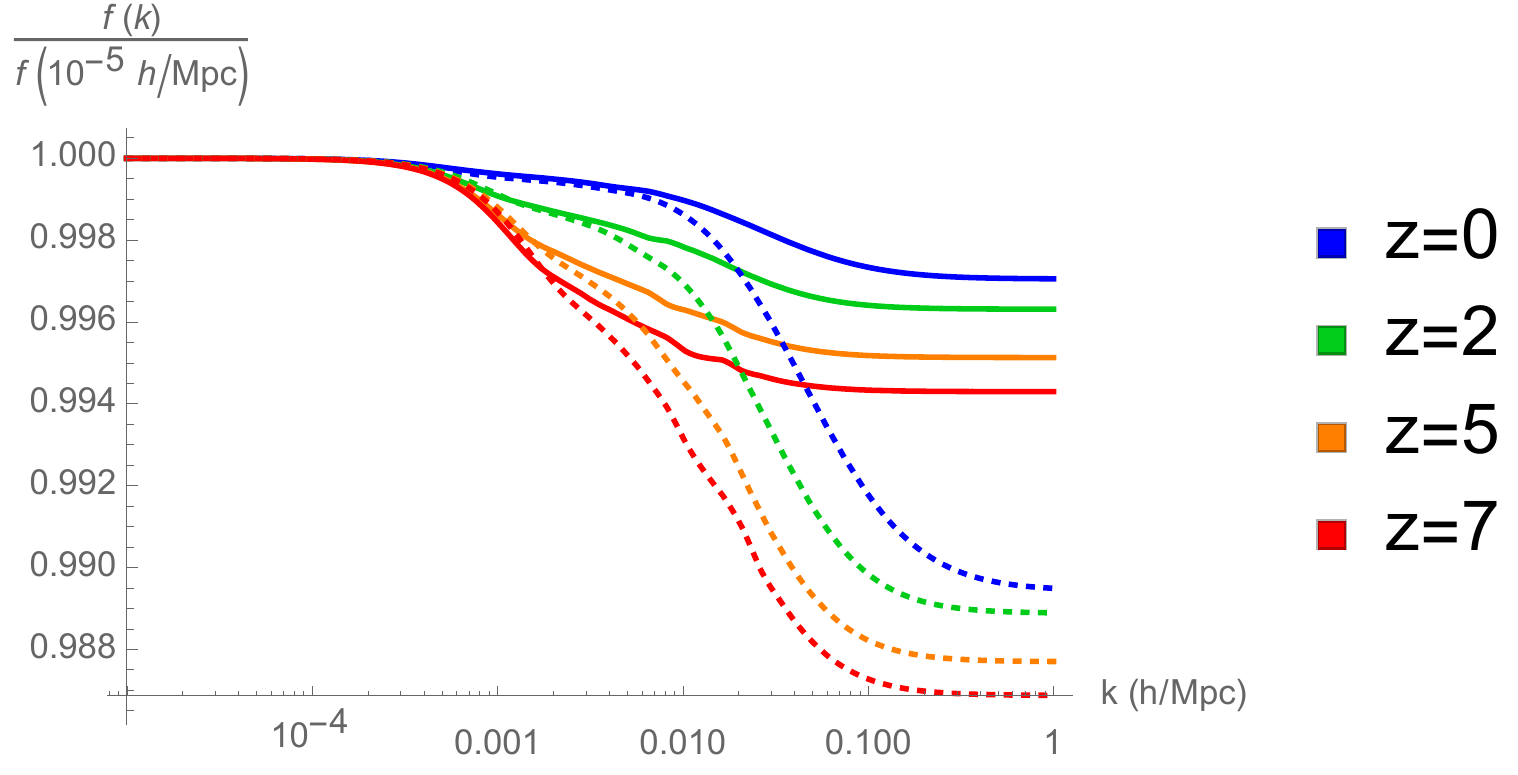}
\caption{
The growth rate of structure $f(z,k)$ as a function of $k$ compared to its low $k$ value 
at $z=0,2,5,7$ for total neutrino mass $m_T=0.058$ eV (solid lines) and $ m_T=0.23$ eV (dotted lines). The values $m_T=0.058$ eV and $ m_T=0.23$ eV correspond to the minimum and maximum 95\% CL constraints on total neutrino mass~\cite{2014PhRvD..89i3018C,Agashe:2014kda}.
}
\label{f(k)z=0-7}
\end{figure}

\section{Sensitivity of cosmological surveys}
\label{surveys}

Ignoring horizon-scale corrections \cite{2012PhRvD..85b3504J} and assuming a simple linear
bias model, the galaxy number density measured in redshift space can be expressed as
\be\label{Pg}
\delta_g(z,\vec{k})=\bigl(b+ f(z,k)\mu^2\bigr) \delta_m(z,k) \  ,
\ee 
where $b$ is the galaxy bias and $\mu\equiv \hat{k}\cdot\hat{r}$, with $\hat{r}$ being the line of sight. 
The galaxy power spectrum $P_g(k,z)$ is then simply related to the matter power spectrum $P_m(k,z)$:
\be\label{Pg-Pm}
P_g(z,\vec{k})=\big[b+ f(z,k)\mu^2\big]^2 P_m(z,k)~,
\ee

Neutrino masses affect not only the mean $f$ at a given redshift but also its weak dependence in $k$. 
Given a neutrino mass hierarchy, we can calculate this $f(z,k)$ from the matter power spectrum. 

Typically $f$ varies between $0.3-1\%$ as we go from low $k$ to high $k$ (see figure~\ref{f(k)z=0-7} )
While the effect is small, we will show
that it is well within the grasp of cosmological surveys that cover large fractions of the sky at redshifts past $z\sim 1$. While measuring $f$ on its own will be a valuable probe of neutrino masses, measuring the scale dependence will be an extremely robust signature.  To study how well one could measure $f(z,k)$, we model $f(z,k)$ 
at a given redshift as being the fiducial relation expected for massive neutrinos within
a given splitting scheme, $f_{0}(z,k)$, 
having corrections that are piecewise constant in two bins in $k$:
\be \label{fbins}
f(z,k) \equiv f_{0}(z,k) + f_L\theta(k_{\rm split}-k)+f_H \theta(k-k_{\rm split}), 
\ee
where $\theta$ is the Heaviside step function, $f_L$ is a possible correction of $f(z,k)$ at low $k$ and $f_H$ is that at high $k$, with the split between high and low $k$ occurring at $k_{\rm split}$.
We take $k_{\rm split}=0.03$h/Mpc based on inspection of figure~\ref{f(k)z=0-7}.
Note that we are not approximating $f(z,k)$ by a step function, only the deviation of $f(z,k)$ from the $f_0(z,k)$ which we calculated using the CLASS Boltzman code~\cite{2011JCAP...07..034B,  2011JCAP...09..032L}.

Given a specified galaxy survey, it is possible to use the galaxy number density to calculate the signal-to-noise expected per mode $k$ and use this to forecast the precision on parameters of interest.
Our intent is to instead characterize the required survey parameters.
To estimate the sensitivity of a survey at a particular redshift, 
we assume that one has a sample-variance limited
measurement of the galaxy number density to a certain limiting $k_{\rm max}$ in harmonic space. 
This highest $k$ is an essential input to a calculation
of constraints on neutrino mass, since there are many more modes at high $k$ than at low $k$. In our calculation we set $k_{\rm max}$ to be the
smaller of either the non-linear scale at a given $z$ or $k_{\rm max}=1$ h/Mpc.

Any useful discussion of
forecast constraints on the total neutrino mass or determinations of inverted versus normal hierarchy must have a reliable accounting of how
small-scale (high $k$) constraints are limited by either non-linearity in the cosmological 
density field, shot noise in the galaxy survey, or details of survey limitations that 
make it difficult to accurately measure density contrasts at small separations. 
However, for the purposes of comparing $f$ on small scales vs large scales, the vastly 
larger number of modes on small scales means that the accuracy
of the comparison between large scales and small scales will primarily be limited by the accuracy of the 
large-scale (low $k$) measurement. 

We work with seven free parameters:
the growth parameters $f_L$ and $f_H$, the sum of the neutrino masses $m_T$, the spectral index $n_s$, the effective number of relativistic neutrinos $N_{\rm eff}$,
the density of cold dark matter $\Omega_{\rm cdm}$ and the galaxy bias $b$. We do this in a minimal 3-flavour neutrino mixing scheme, with all the other cosmological parameters fixed. We fix the amplitude of the matter power spectrum, as any change in the matter power
spectrum amplitude could be absorbed into a corresponding correlated change in $b, f_L$ and $f_H$. 
In particular we take 
$h=0.7, T_{\gamma0}=2.7255~{\rm K}, \Omega_b=0.05,\Omega_\Lambda=1-\Omega_m, \sigma_8=0.83$.

We take the spectral index $n_s=0.96$ with a prior given by the Planck 2015 \cite{2015arXiv150201589P} 68\%~C.L. value of $\sigma_{n_s}=0.006$. 
We take the Standard model value for the effective number of relativistic neutrino degrees of freedom in the early universe, $N_{\rm eff}=3.046$ with a prior of $0.34$ \cite{Agashe:2014kda}.\footnote{To obtain $N_{\rm eff}=3.046$ using CLASS Boltzman code with three massive neutrinos we need to set the ultra-relativistic species number $N_{ur}=0.00641$. This is documented in the explanatory.ini file of the publically available version 2.4.3, http://class-code.net/.}
We take $\Omega_{\rm cdm}=0.25$ with a prior of $0.01$. We take the bias $b=1$ with no prior. For $m_T$ we consider all neutrino masses that satisfy the current 95\% CL constraints~\cite{2014PhRvD..89i3018C,Agashe:2014kda}. The corresponding range of neutrino masses is [0.058, 0.23] eV for the normal hierarchy (NH) and [0.098, 0.23] eV for the inverted hierarchy (IH). 

Given an assumed galaxy number density field up to some $k_{\rm max}$, we can calculate the information content of an assumed survey using
the Fisher matrix: the ensemble-averaged curvature with respect to parameters of the logarithm of the likelihood function for an assumed
fiducial set of input parameters \cite{10.2307/2342435}. By assuming a sample-variance-limited survey, at each wavenumber $k$ there is no noise and the precision
of the power measurement within the survey is solely limited in precision by the number of modes that are measured.

The Fisher matrix entries ${\bf F}_{x,y}$ are given by (\cite{1997ApJ...480...22T},\cite{1997PhRvL..79.3806T},\cite{Seo:2003aa})
\be\label{fisherdef}
{\bf F}_{x,y}\approx{1\over2}\int {dk^3\over(2\pi)^3} ~ [\partial_x \ln P_g(z,\vec{k})] ~ [\partial_y \ln P_g(z,\vec{k})] ~V_{\rm eff}(\vec{k})
\ee
$V_{\rm eff}$ is the effective volume of the survey. For a constant comoving galaxy number density $n_g$ and galaxy survey volume "Vol" the effective volume is given by:
\be\label{Veff}
V_{\rm eff}(z,\vec{k})= {\rm Vol}~\Bigg[{  1\over 1+1/(n_g P_g(z,\vec{k})) }\Bigg]^2
\ee
The $1/n_g$ is the white shot noise from the Poisson sampling of the density field. When $s\equiv  1/(n_g b^2 P_m) $ is much less than one, shot noise is negligible. 
The indices $x,y$ refer to the parameters $f_L, f_H, m_T, n_s, N_{\rm eff}, \Omega_{\rm cdm},b$.
%
%
\def\shotcorrection{\Bigg[{  1\over 1+{s\over (1+f\mu_{\hat{k}}^2/b)^2} }\Bigg]^2}
For $x,y\in \{f_L, f_H, m_T, n_s, N_{\rm eff}, \Omega_{\rm cdm}\}$ this leads to:
\be
\label{fisher}
{\bf F}_{x,y} = {{\rm Vol}\over2}\int {dk^3\over(2\pi)^3}
\Bigg[ {2\mu_{\hat{k}}^2 \partial_x f/b\over 1+f\mu_{\hat{k}}^2/b} + \partial_x \ln P_m\Bigg]
\Bigg[ {2\mu_{\hat{k}}^2 \partial_y f/b\over 1+f\mu_{\hat{k}}^2/b} + \partial_y \ln P_m\Bigg]
\shotcorrection
\ee
Note that $\partial_{f_L} [\ln P_m]=\partial_{f_H} [\ln P_m]=0$.

For $x\in \{f_L, f_H, m_T, n_s, N_{\rm eff}, \Omega_{\rm cdm}\}$, $y=b$:
\be
{\bf F}_{x,b} = {{\rm Vol}\over2}\int {dk^3\over(2\pi)^3}
\Bigg[ {2\mu_{\hat{k}}^2 \partial_x f/b\over 1+f\mu_{\hat{k}}^2/b} + \partial_x \ln P_m\Bigg]
\Bigg[ {2/b\over 1+f\mu_{\hat{k}}^2/b} \Bigg]
\shotcorrection
\ee
and for $x=y=b$
\be
{\bf F}_{b,b} = {{\rm Vol}\over2}\int {dk^3\over(2\pi)^3}
\Bigg[ {2/b\over 1+f\mu_{\hat{k}}^2/b} \Bigg]^2
\shotcorrection
\ee
The integrals are evaluated in two parts. Since $\int {dk^3\over(2\pi)^3}={1\over 4\pi^2} \int_{0}^{k_{\rm max}} k^2 dk \int_{-1}^1 d\mu_{\hat{k}}$, we evaluate the $d\mu_{\hat{k}}$ integral analytically with Mathematica. We then evaluate the $dk$ integral numerically.  
The integral over $dk$ is cut off at $k_{\rm max}$, assumed to be set by the
smaller of either the non-linear scale at a given $z$ or by $k_{\rm max}=1$ h/Mpc. We define the nonlinear scale at a given redshift  to be the point where the linear power spectrum and the halofit model ~\cite{Smith:2002dz,2012MNRAS.420.2551B,2012ApJ...761..152T}
differ by 10\%. 
The particular cutoff
assumed for $k_{\rm max}$ does not materially change our conclusions, as the ability to measure the
scale dependence of $f(k,z)$ is primarily limited by the number of available low-$k$ modes.
To the ${\bf F}^{}_{x,x}$ entries, $x = n_s, N_{\rm eff}, \Omega_{\rm cdm}$
we add $1/\sigma_{x}^2$ corresponding to the Planck 2015 68\%~C.L. on these $x$, as discussed above.
We evaluate $P_m(z,k)$ and $f(z,k)$ using the CLASS Boltzmann code \cite{2011JCAP...07..034B,  2011JCAP...09..032L}.
 
The one sigma uncertainty in a parameter $x$ is given by the square root of the of the diagonal elements of the inverse of the Fisher matrix, $\sigma_x=\sqrt{ {\bf F}^{-1}_{x,x} }$. We are interested in $x=f_L$ and $f_H$. 

\section{Results and Discussion}
\label{results}

\begin{figure}
\centering
\begin{subfigure}[b]{0.45\textwidth}
\includegraphics[width=\textwidth]{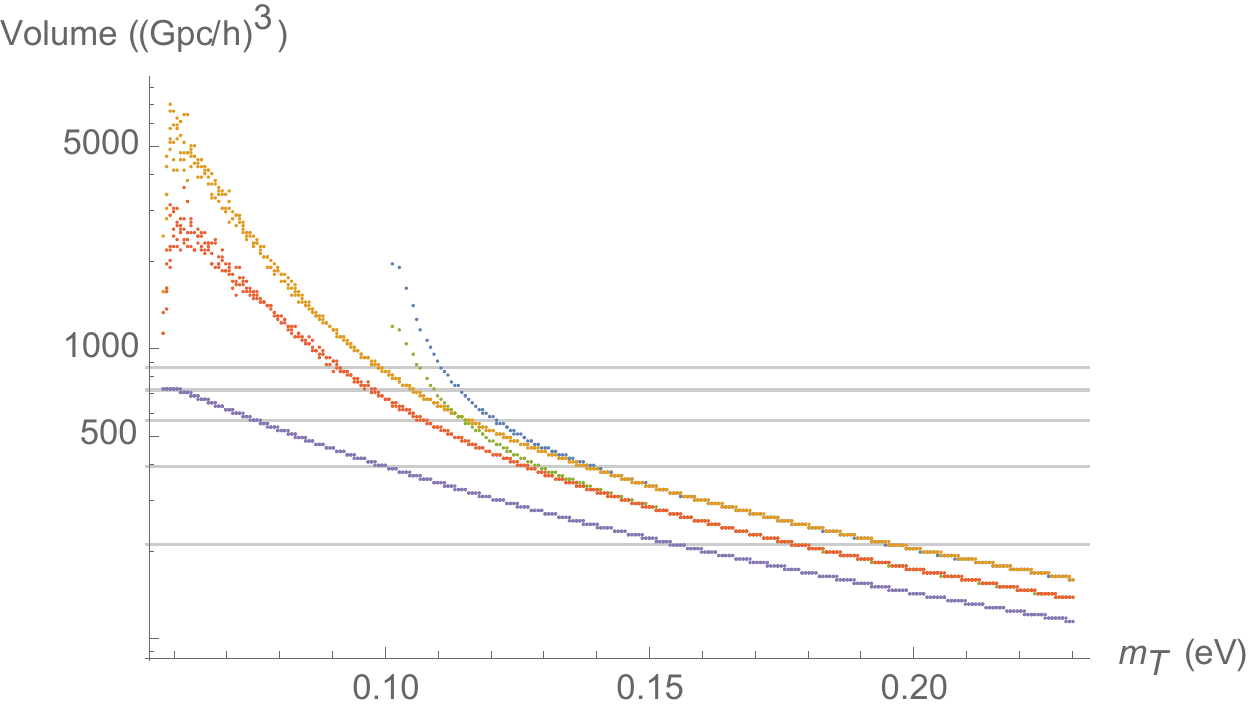}
\caption{$n_g^{-1}=0~({\rm Mpc}/h)^3$}
\end{subfigure}
\begin{subfigure}[b]{0.45\textwidth}
\includegraphics[width=\textwidth]{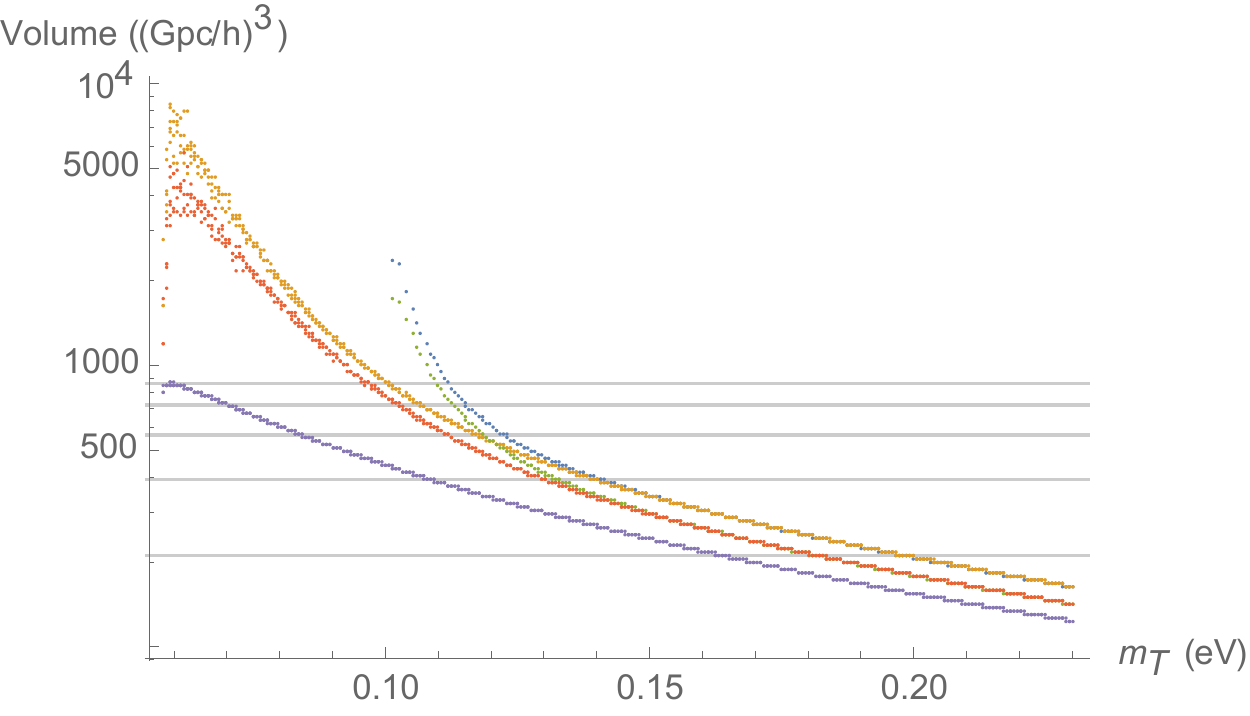}
\caption{$n_g^{-1}=100~({\rm Mpc}/h)^3$}
\label{VolfLHb}
\end{subfigure}
\begin{subfigure}[b]{0.4\textwidth}
\includegraphics[width=\textwidth]{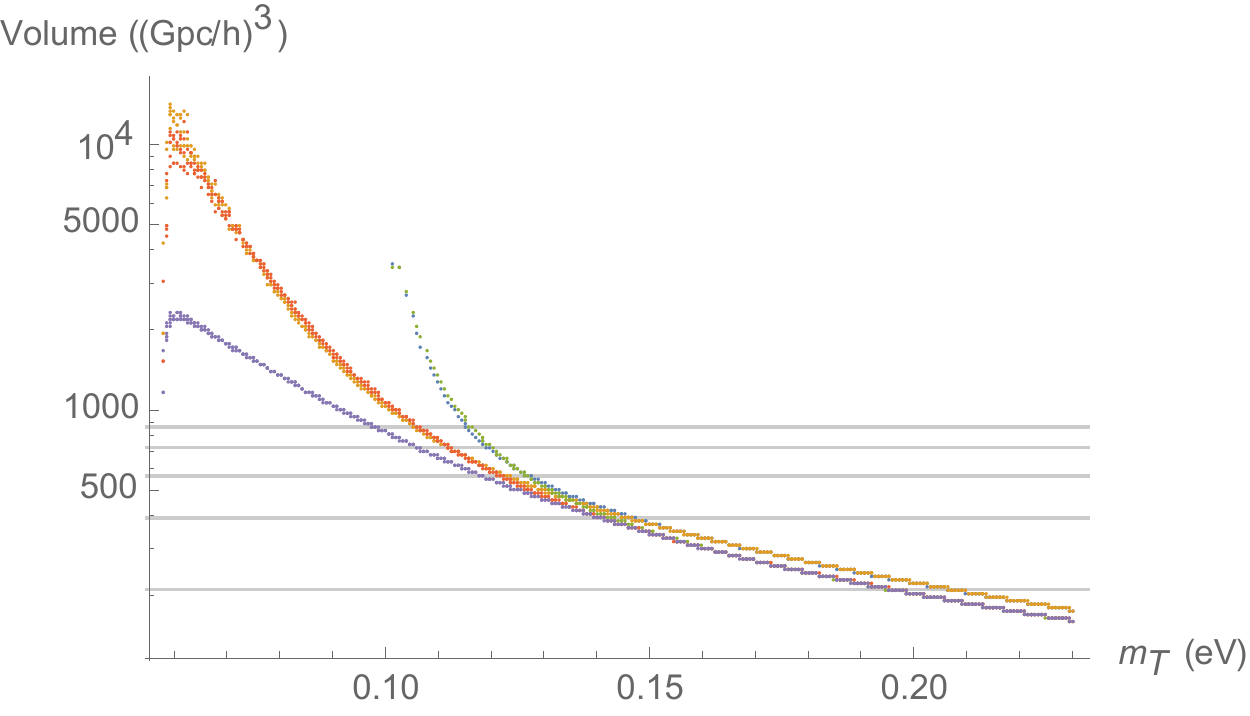}
\caption{$n_g^{-1}=500~({\rm Mpc}/h)^3$}
\end{subfigure}
\begin{subfigure}[b]{0.59\textwidth}
\includegraphics[width=\textwidth]{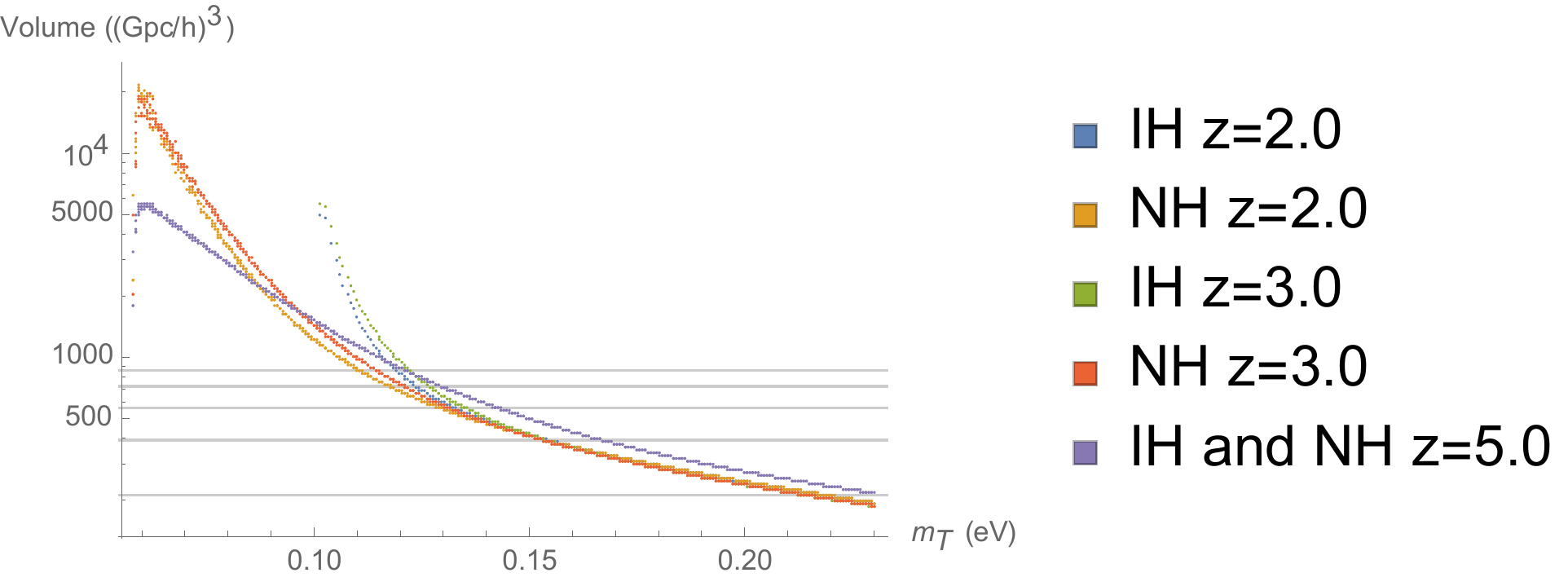}
\caption{$n_g^{-1}=1000~({\rm Mpc}/h)^3$}
\label{VolfLHd}
\end{subfigure}
\caption{
Volume needed, for different values of the comoving galaxy number density $n_g$, 
for a $1 \sigma$  determination of  ${\sigma_{f_L}\over f(k_{\rm low})-f(k_{\rm high})}$ as a function of total neutrino mass $m_T$ for the NH and IH. The IH $z=5$ line is only for $m_T > 0.1$ eV. 
The horizontal grey lines corresponds to the volume available up to redshifts 2,3,4,5 and 6. }
\label{VolfLH}
\end{figure}

By inverting the Fisher matrix we are able to calculate $\sigma_{f_L}$ and $\sigma_{f_H}$. As expected, the vastly greater number of modes at high $k$ results in $\sigma_{f_H}$ being an order of magnitude smaller than $\sigma_{f_L}$. Hence measuring the scale dependence of $f(k)$ requires measuring $f_L$ to within the percentage difference between low and high $k$ shown in figure~\ref{f(k)z=0-7}. 
Rather than present the ratio of 
$\sigma_{f_L}/(f(k_{\rm low})-f(k_{\rm high}))$, we show the volume needed to measure this ratio to ($1\sigma$) in figure~\ref{VolfLH}.  

For both the inverted and normal hierarchies, a variety of redshifts, and different values of the comoving galaxy number density, we show in figure~\ref{VolfLH} the volume required to 
minimally measure at $1\sigma$ the expected difference between the low-$k$ and high-$k$ growth rates. These constraints are driven primarily by volume, rather than by the smallest resolvable scales. 

%


The volumes required are large compared to ongoing surveys.
While challenging for a traditional galaxy survey, this is not
an unattainable goal for emission-line-galaxy surveys or intensity mapping experiments. Rather, the challenge is to reach sufficient number densities or sufficiently low noise levels to image large-scale structure at the required sensitivity.  For example, the LoFar 21 cm intensity mapping experiments will survey half the sky between redshifts of $z=1.5$ to $z=10$, corresponding to volumes of over 1100~Gph/h$^3$~\cite{vanHaarlem:2013dsa}. 
From figure~\ref{VolfLHb} we see that for $n_g > 10^{-2} h^3$/Mpc$^{-3}$ a redshift survey out to $z=5$ should measure at $1\sigma$ this scale dependence for all neutrino masses, whereas a survey out to $z=3$ would measure at $1\sigma$ this scale dependence if the total neutrino mass was greater than 0.12 eV. 
In particular a volume of about 900~Gph/h$^3$ is needed to measure the scale dependent growth rate for all possible neutrino masses at around $z=5$.
For $n_g$ is as low as $10^{-3} h^3$/Mpc$^{-3}$, we see from figure~\ref{VolfLHd} that a survey out to $z=3$ would measure at $1\sigma$ this scale dependence if the total neutrino mass was greater than 0.14 eV. 

The measurement of the scale dependence of $f(z,k)$ is a cosmological measurement of neutrino masses, complementary to their measurement by oscillatory experiments. And since the amplitude and shape of $f(z,k)$ depends on the particular set of neutrino masses, its measurement not only determines the total neutrino mass, but could also distinguish between the IH and NH mass hierarchies. 

\acknowledgments
The work grew out of fruitful discussion with Gil Holder and Gabrielle Simard. I am grateful to Jim Cline for helpful comments in preparing the final draft of this manuscript. 
OH was supported by the FQRNT Programme de recherche 
pour les enseignants de coll\`ege.
This research was enabled in part by support provided by Calcul Qu\'ebec (www.calculquebec.ca) and Compute Canada (www.computecanada.ca).

\bibliography{neuMassRSDv9_final}

\end{document}